# Impact of Distinct $Ca^{2+}$ Sources on the Physical Properties of Alumina-based Macroporous Refractories for Thermal Insulation at High Temperature

O. H. Borges[1]*, T. Santos Jr[1], R. R. B. de Oliveira[1], V. R. Salvini[2] and V. C. Pandolfelli[1]

[1] Materials Microstructure Engineering Group (GEMM), FIRE Associate Laboratory, Materials Engineering Department and Graduate Program in Materials Science and Engineering, Federal University of São Carlos (UFSCar), Brazil.
[2] College of Technology (FATEC), Sertãozinho, Brazil.

**Abstract**
Calcium aluminate cement (CAC), calcium carbonate ($CaCO_3$), calcium hydroxide [$Ca(OH)_2$] and calcium oxide (CaO) were investigated in alumina-based macroporous compositions for *in situ* formation of hibonite (also known as $CaO \cdot 6Al_2O_3$ or $CA_6$). Due to its volumetric expansion, this phase could counteract the linear shrinkage observed when macroporous ceramics are fired. In order to evaluate the impact of each $Ca^{2+}$ source on the physical properties, different formulations were processed and characterized for their total porosity, crushing strength and linear shrinkage. $CaCO_3$-containing samples presented high porosity and small dimensional changes after thermal treatment. Thus, a novel composition was formulated with $CaCO_3$ in order to result in 100 % $CA_6$ on thermodynamic equilibrium. Besides the usual properties, its *in situ* changes in the hot elastic modulus ($E_{in\ situ}$) and thermal conductivity ($k_{eff}$), were evaluated. This composition presented promising results, as a constant and low $k_{eff}$, low linear expansion and Young modulus increase at low temperature. These properties make it possible to produce more efficient insulators with better performance in service.

## 1. Introduction

Global energy demand has been growing steadily with the time. According to the World Energy Outlook 2018, published by International Energy Agency (IEA), it will increase by 25 % until 2040. By this scenario, industries play a key role, as 37 % of all energy consumed in the world in 2016 was used in this sector. Additionally, about one-third of this energy was used for high temperature processes (>1000 ºC)[1].

Aiming to reduce energy losses related to heat transfer, refractory ceramic fibers (RCF) are the main materials used for high temperature insulation linings[2]. Although they present low thermal conductivity and some might be toxic when inhaled[3], these materials experience densification in service, which causes an increase in their $k_{eff}$. Macroporous refractory ceramics are pointed out as a safer and better alternative for RCF, as they are non-toxic, heat-resistant and could be dimensionally stable at high temperatures. By definition, they are comprised by a refractory matrix containing at least 40 % of their volume formed by pores with a diameter greater than 50 nm.

Considering the high amount of pores dispersed in their microstructures, macroporous ceramics undergo high densification and undesired dimensional changes when fired. The *in situ* formation of expansive phases, *e.g.* $CA_6$, could then counteract the samples shrinkage. When $CA_6$ is formed by the reaction between alumina and calcium sources, a volumetric expansion close to 16 % is observed[4]. Additionally, $CA_6$ is a refractory phase with low effective thermal conductivity and high chemical stability up to 1600 ºC. Due to these properties, macroporous ceramics containing hibonite are attractive materials for thermal insulation at high temperature. Based on that, the production of macroporous refractory ceramics containing alumina and calcium sources were investigated in this work. Their physical properties (porosity, linear shrinkage and mechanical strength) were evaluated and the results are discussed considering their application as thermal insulators up to 1600 ºC.

## 2. Experimental Procedure
### 2.1 Materials

Refractory ceramics were prepared by a direct foaming method consisting on the preparation of an alumina suspension and a liquid foam stabilized by a surfactant. Their compositions are listed in **Table 1**. The liquid foam was incorporated in the alumina suspension, resulting in a foamed suspension.

**Table 2** shows the $Ca^{2+}$ sources used for each formulation and the $CA_6$ amount expected to be formed under thermodynamic equilibrium of these systems after thermal treatment. $CaCO_3$ (RHI Magnesita, Brazil) and $Ca(OH)_2$ (Synth, Brazil) were added to the $Al_2O_3$ suspension before foam incorporation. Conversely, calcium aluminate cement (Secar 71, Imerys Aluminates, France) and calcium oxide (obtained after calcining $CaCO_3$ at 900 ºC for 4 hours) were added at the last processing step due to their accelerated hydration kinetics.

For the reference composition (REF), 1 wt % of hydratable alumina (Alphabond 300, Almatis, Germany) was used as binder. Due to the presence of CaO as impurity, a minor amount of $CA_6$ was expected to be formed in the REF samples.

**Table 1 Alumina liquid foam composition**

| Ceramic foam composition | | |
|---|---|---|
| | Components | wt % |
| $Al_2O_3$ Suspension | Almatis CL370 | 76.06 |
| | Almatis CT3000SG | 5.72 |
| | BASF Castament FS60 | 0.09 |
| | BASF Lutensol AT50 | 0.10 |
| | Distilled Water | 15.79 |
| Foam | BASF Vinapor GYP 2680 | 3.47 |
| | Down Cellosize 100 CG FF | 0.0056 |

**Table 2 Binders and lime sources used for each ceramic foam formulation**

| Formulations | Additives (Wt %) | | | | Expected $CA_6$* (wt %) |
|---|---|---|---|---|---|
| | CAC | $CaCO_3$ | $Ca(OH)_2$ | CaO | |
| REF | - | - | - | - | <0.5 |
| CAC | 5.00 | - | - | - | 20 |
| $CaCO_3$/CAC | 1.00 | 2.04 | - | - | 20 |
| $Ca(OH)_2$/CAC | 1.00 | - | 1.42 | | 20 |
| CaO | - | - | - | 1.55 | 20 |
| (+)$CaCO_3$/CAC | 1.00 | 12.90 | - | - | 100 |

\* Under thermodynamic equilibrium.

The foamed ceramic suspensions were cast into 50 mm x 50 mm cylinders, 150 mm x 25 mm x 25 mm bars and 230 mm x 114 mm x 64 mm bricks. All samples were cured at 50 ºC for 24 h. Besides the temperature control, the formulations containing calcium aluminate cement were cured under humidity of 80 %.

After that, all samples were dried at 110 ºC for 24 h. Finally, some samples of each formulations were fired at 1600 ºC for 5 hours.

**2.2 Methods**

Linear shrinkage of samples was evaluated by their diameter measurements before and after thermal treatment at 1600 ºC. Total porosity of each formulation was obtained by the ratio between macroporous samples volumetric density (measured by immersion in water) and the density of their solid fraction [measured for grounded samples using helium picnometry (AccuPyc 1330, Micromeritics, USA)]. Crushing strength measurements were carried out according to ASTM 133-97 in an universal testing machine (MTS 810, MTS, Eden Prairie, USA) with a load cell of 50 kN.

In addition, REF e (+)$CaCO_3$/CAC had their E$_{in}$ $_{situ}$ versus temperature profile evaluated by sonic resonance technique (ScanElastic 02, ATCP Physical Eng., Brazil) according to ASTM C1875-00. Young modulus was measured every 5 minutes, up to 1400 ºC and back to room temperature, with heating and cooling rate of 2 ºC·min$^{-1}$.

Thermal conductivity of (+)$CaCO_3$/CAC was measured by parallel hot wire method using a TCT426 (Netzsch, Germany), according to ASTM C1113. The test was carried out from room temperature to 1200 °C with 200 °C steps.

**3. Results and discussion**
**3.1 Different $Ca^{2+}$ sources comparison**

**Fig.1** shows the linear shrinkage and total porosity of green and fired samples. Green samples attained a high porosity (> 75 %vol). However, after firing at 1600 ºC a lower value could be observed, which is related to their densification. Composition REF showed the lowest porosity after firing, which is associated to its linear shrinkage. Considering that this composition did not contain additional $Ca^{2+}$ sources, the effect of the *in situ* $CA_6$ formation was not observed. Therefore, the $CA_6$ formation counteracted linear shrinkage in the compositions containing lime sources. Additionally, considering that hibonite formed after thermal treatment can grow in equiaxial or acicular morphology[5], it's believed that lower linear shrinkage values can also be related to a higher $CA_6$ aspect ratio. The generation of acicular hibonite implies in intrinsic porosity generation, which leads to further expansion.

**Table 3** presents the mechanical strength of green and fired compositions. When formulations containing $Ca^{2+}$ sources are compared, those with lower linear dimensional changes also had higher crushing strength after fired. This effect could be explained by the formation of higher aspect ratio $CA_6$, which can induce strengthening mechanisms, improving material's mechanical properties. Additionally, porosity played a key role in mechanical strength, as highlighted by REF composition. Its porosity is almost 20 % lower than the average value recorded for compositions containing calcium sources. Consequently, its crushing strength reached 62 MPa, whereas other compositions attained up to 7.4 MPa.

On the other hand, green crushing strength mainly depends on binder performance. In this regard, Alphabond used in REF samples lid to the highest value observed. Combination of CAC and $CaCO_3$ resulted in high mechanical strength, what is most likely due to the dispersant effect that this carbonate induces to CAC[6]. Adding calcium oxide as a binder showed results comparable to those presented by CAC without additives or with

Ca(OH)$_2$.

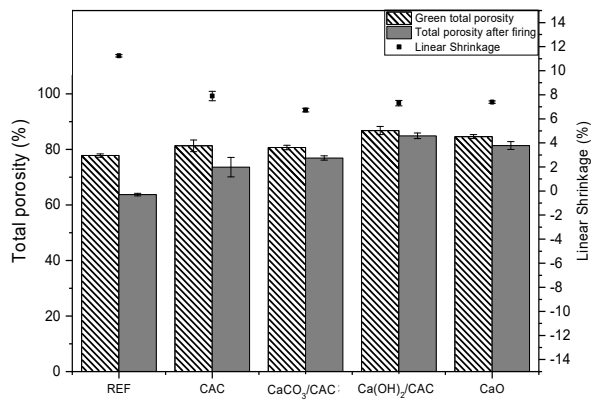

**Fig. 1** Linear shrinkage and total porosity evaluated before and after firing the samples at 1600 ºC.

**Table 3** Composition's mechanical strength in their green stage and fired at 1600 ºC for 5 hours

| Composition | Green crushing strength (MPa) | Crushing strength after firing (MPa) |
|---|---|---|
| REF | 0.95 ± 0.03 | 62 ± 2 |
| CAC | 0.065 ± 0.006 | 2.2 ± 0.2 |
| CaCO$_3$/CAC | 0.51 ± 0.04 | 7.4 ± 0.4 |
| Ca(OH)$_2$/CAC | 0.060 ± 0.004 | 4.2 ± 0.4 |
| CaO | 0.096 ± 0.007 | 4.2 ± 0.5 |

Based on the results discussed, CaCO$_3$/CAC was the most promising formulation due to its lower linear shrinkage and a good balance between total porosity and crushing strength after firing. Therefore, another macroporous refractory composition containing a higher content of CaCO$_3$ (designed to present 100 % of CA$_6$ after thermal treatment) was prepared. This formulation, named (+)CaCO$_3$/CAC, was processed following the same procedures presented before and their characterization results are discussed below.

### 3.2 Al$_2$O$_3$-CaCO$_3$ macroporous ceramics designed to present 100 % CA$_6$

**Table 4** shows the (+)CaCO$_3$/CAC linear shrinkage, total porosity and crushing strength results. Although their values presented some changes when compared to CaCO$_3$/CAC, the linear shrinkage one stands out. After thermal treatment, (+)CaCO$_3$/CAC linearly expanded around 0.81 %. In other words, the expansion due to CA$_6$ formation exceeded the shrinkage and resulted in almost 1 % of unidimensional growth.

**Table 4** Linear shrinkage, total porosity and crushing strength of (+)CaCO$_3$/CAC.

| Linear shrinkage | Fired Total porosity | Green Crushing strength | Crushing strength after firing |
|---|---|---|---|
| (-0.81±0.09) % | (80±1) % | (0.60±0.10) MPa | (6.2±0.6) MPa |

Besides this expansion effect, it was reported that the use of CaCO$_3$ in alumina-based dense formulations results in earlier increase of their Young's modulus E[7]. Thus, aiming to evaluate if this phenomenon could take place in macroporous systems, (+)CaCO$_3$/CAC and REF compositions had their E$_{in\ situ}$ measured. Elastic modulus evolution with the temperature is presented in **Fig.2**.

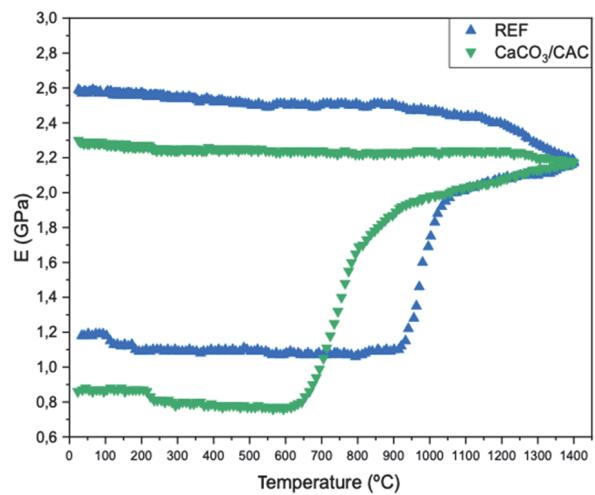

**Fig.2** Elastic modulus evolution with the temperature for REF and (+)CaCO$_3$/CAC.

REF composition presented higher initial and final Young modulus – what was expected because of its higher crushing strength at green stage and higher densification after firing. However, REF presented E initial increase just at 1000 ºC whereas for composition (+)CaCO$_3$/CAC it was 650 ºC.

The advantages on the earlier E increase are: *(i)* these macroporous insulators could be thermally treated at a lower temperatures for transportation and installation, finishing *in situ* their firing process and *(ii)* using lower sintering temperatures to produce macroporous ceramics that should be applied in low thermal demand environments, *e.g.* aluminum industry. In either cases, the energy input required to manufacture the macroporous material would be reduced.

Nevertheless, despite the importance of reducing linear shrinkage and strengthening temperature of macroporous ceramics, the key property for this material is the thermal conductivity. Thus, (+)CaCO$_3$/CAC thermal conductivity (k$_{eff}$) was measured up to 1200 °C, as shown in **Fig. 3**, where it is compared to values of a commercial alumino-silicate fiber according to reference 8).

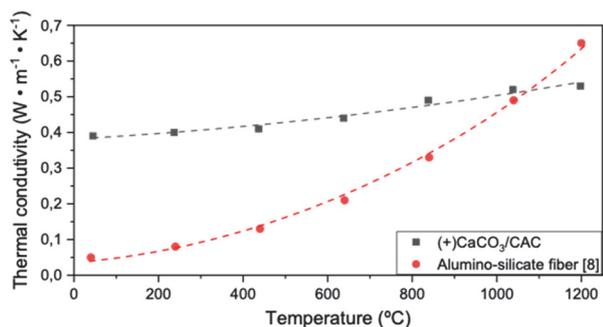

**Fig.3 Thermal conductivity of (+)CaCO$_3$/CAC and other insulating materials.**

(+)CaCO$_3$/CAC showed $k_{eff}$ value at room temperature below 0.4 W·m$^{-1}$K$^{-1}$. This low $k_{eff}$ was achieved mainly by its high porosity and CA$_6$ presence in the microstructure, which is an intrinsic low thermal conductivity phase[9]. However, at room temperature the commercial fiber had better thermal insulation performance than (+)CaCO$_3$/CAC.

On the other hand, it is known that fiber-based insulators usually have a significant $k_{eff}$ increase with temperature. This behavior is due to its microstructure that is ineffective to halt heat transfer by thermal radiation, which is the main thermal transfer phenomenon at high temperature. Therefore, (+)CaCO$_3$/CAC, whose pores are able to spread thermal radiation, presented lower $k_{eff}$ above 1050 ºC. Thus, this refractory macroporous ceramic should show better performance for thermal insulation at high temperatures, when compared to refractory fiber modules.

## 4. Conclusions

The effects of different Ca$^{2+}$ sources in alumina-based macroporous ceramics were studied. All formulations resulted in green porosities above 75 vol % and after thermal treatment they decreased according the linear shrinkage. Compositions with expected CA$_6$ formation of 20 wt % showed linear shrinkage between 6 % and 8 %, instead of 12 % presented by REF composition. In this regard, CaCO$_3$/CAC composition showed the lowest value, most likely due to its capacity to form CA$_6$ with higher ratio aspect. These aspects highlighted the impact of CA$_6$ formation on the linear shrinkage. Additionally, the higher crushing strength presented by CACO$_3$/CAC composition after firing is another evidence of acicular CA$_6$ formation.

Because of its promising results, a composition containing higher content of CaCO$_3$ resulting in 100 % of CA$_6$ formation was prepared and named (+)CaCO$_3$/CAC. This composition showed a linear expansion (+0.81 %), mitigating the pronounced volumetric changes that this type of material undergoes after firing. Additionally, its Young's modulus increasing started at ~650 ºC, much lower than for the REF (~1000 ºC). This aspect can enable the reduction of energy required to manufacture macroporous insulators.

Finally, (+)CaCO$_3$/CAC presented low thermal conductivity values at temperature range between 30 ºC and 1200 ºC. In contrast to commercial fiber insulating, it was worth observing that its $k_{eff}$ changed a little with temperature. Because of that, (+)CaCO$_3$/CAC started to be a more efficient thermal insulator than commercial alumino-silicate fiber above 1050 ºC.

## 5. Acknowledgments

This study was financed in part by Coordenação de Aperfeiçoamento de Pessoal de Nível Superior - Brasil (CAPES) - finance code 001, by Conselho Nacional de Desenvolvimento Científico e Tecnológico – Brasil (CNPq) – process number 130843/2018-0 and by Fundação de Amparo à Pesquisa do Estado de São Paulo – Brasil (FAPESP) – process number 2018/07745-5.

Additionally, the authors are thankful for the support of BASF, RHI Magnesita, and FIRE (Federation for International Refractory Research and Education).